%% file: 0-Main.tex
\documentclass[
twocolumn,
]{ceurart}

\input{preamble/packages}

\begin{document}

\copyrightyear{2024}
\copyrightclause{Copyright for this paper by its authors.
  Use permitted under Creative Commons License Attribution 4.0
  International (CC BY 4.0).}

\conference{The 1st Workshop on Risks, Opportunities, and Evaluation of Generative Models in Recommender Systems (ROEGEN@RecSys 2024), October 2024, Bari, Italy.}

\title{Cognitive Biases in Large Language Models for\\News Recommendation}


\author[1]{Yougang Lyu}[%
orcid=0009-0000-1082-9267,
email=youanglyu@gmail.com,
]
\address[1]{University of Amsterdam, Amsterdam, The Netherlands}

\author[2]{Xiaoyu Zhang}[%
orcid=0000-0002-5667-1036,
email=tinyoctopus1999@gmail.com,
]
\address[2]{Shandong University, Qingdao, China}

\author[3]{Zhaochun Ren}[%
orcid=0000-0002-9076-6565,
email=z.ren@liacs.leidenuniv.nl,]
\address[3]{Leiden University, Amsterdam, The Netherlands}

\author[1]{Maarten	{de Rijke}}[%
orcid=0000-0002-1086-0202,
email=m.derijke@uva.nl,
]

\input{sections/00-abs}


\maketitle

\input{sections/00-background}
\input{sections/01-influence}
\input{sections/02-mitigate}
\input{sections/03-related_work}
\input{sections/04-conclusion}

\begin{acknowledgments}
    This research was (partially) supported by the Dutch Research Council (NWO), under project numbers 024.004.022, NWA.1389.20.\-183, and KICH3.LTP.20.006, and the European Union's Horizon Europe program under grant agreement No 101070212.
    All content represents the opinion of the authors, which is not necessarily shared or endorsed by their respective employers and/or sponsors.
\end{acknowledgments}
\bibliography{references}




\end{document}

%% file: preamble/packages.tex
\usepackage{listings}
\usepackage{enumitem}

%% file: sections/00-abs.tex
\begin{abstract}
  Despite large language models (LLMs) increasingly becoming important components of news recommender systems, employing LLMs in such systems introduces new risks, such as the influence of cognitive biases in LLMs. Cognitive biases refer to systematic patterns of deviation from norms or rationality in the judgment process, which can result in inaccurate outputs from LLMs, thus threatening the reliability of news recommender systems. Specifically, LLM-based news recommender systems affected by cognitive biases could lead to the propagation of misinformation, reinforcement of stereotypes, and the formation of echo chambers. In this paper, we explore the potential impact of multiple cognitive biases on LLM-based news recommender systems, including anchoring bias, framing bias, status quo bias and group attribution bias. Furthermore, to facilitate future research at improving the reliability of LLM-based news recommender systems, we discuss strategies to mitigate these biases through data augmentation, prompt engineering and learning algorithms aspects.
\end{abstract}

%% file: sections/00-background.tex
\section{Background}

Large language models (LLMs) are becoming crucial components of recommender systems~\cite{DBLP:conf/coling/LiZLC24,DBLP:journals/corr/abs-2305-19860,DBLP:journals/corr/abs-2402-18590,zhang2024towards}. In particular, news recommender systems rely heavily on LLMs to analyze vast amounts of textual data, ensuring that users receive news articles that align with their interests and preferences~\cite{DBLP:conf/sigir/ZhangW23,DBLP:conf/wsdm/LiuCS024,DBLP:conf/ecir/LiZM24}.

Despite their growing importance and effectiveness, the deployment of LLMs in news recommender systems is not without risks~\cite{DBLP:conf/recsys/ZhangBZWF023,DBLP:conf/recsys/HuaLXCZ23,DBLP:conf/www/ZhangB0WF024,jiang2024item,deldjoo2024understanding}. One significant issue that has emerged is the influence of cognitive biases in LLMs on the ability to make decisions correctly~\cite{jones2022capturing,schramowski2022large,DBLP:journals/corr/abs-2311-04076,DBLP:journals/corr/abs-2403-00811,schramowski2022large}. Cognitive biases are systematic patterns of deviation from norm or rationality in judgment, can lead to the production of inaccurate or skewed outputs~\cite{tversky1974judgment,wilke2012cognitive}. Specifically,~\citet{DBLP:journals/corr/abs-2308-00225} find that LLMs fine-tuned on human-generated data are significantly affected by cognitive biases during the inference phase, which seriously affects the reliability of LLM-based news recommender systems.

Identifying and addressing the potential impact of cognitive biases in LLMs for recommender systems is critical, especially in the high-stake news recommendation task. News recommender systems play a crucial role in shaping public opinion, informing decision-making, and influencing societal discourse~\cite{helberger2021democratic}. Therefore, any distortion in the news recommendation process caused by cognitive biases can have far-reaching consequences, potentially spreading misinformation~\cite{spina2023human,lewandowsky2012misinformation}, reinforcing stereotypes~\cite{yunkaporta2023right}, or contributing to echo chambers~\cite{wang2020public}.

Consequently, there is a pressing need to discuss the potential risks of cognitive bias in LLMs for news recommender systems and to explore possible solutions. This leads to our two central research questions: (1) \textit{How do cognitive biases influence the decision-making processes of LLMs for news recommendations?} (2) \textit{What strategies can be employed to mitigate these cognitive biases?} To this end, we analyze the impact of various cognitive biases on LLM-based news recommender systems. Additionally, we discuss strategies for mitigating these biases through data augmentation, prompt engineering, and learning algorithms.

%% file: sections/01-influence.tex
\section{Risks of Cognitive Biases}

News recommender systems aim to personalize the news articles presented to users, tailoring the recommendation based on individual preferences, behavior, and historical data. LLM-based news recommender systems often leverage LLMs to filter and rank news articles, determining which articles to prioritize. However, LLMs trained on large amounts of human-generated data may inherit human cognitive biases, reflecting similar patterns in their outputs~\cite{DBLP:journals/corr/abs-2403-00811,DBLP:journals/corr/abs-2308-00225}. Cognitive biases in LLMs can profoundly influence LLM-based news recommender systems, potentially leading to several adverse effects:
\begin{itemize}[leftmargin=*,nosep]
\item \textbf{Anchoring bias~\cite{tversky1974judgment}:} The reliance on the first piece of information (the ``anchor'') when making decisions. Assuming a conversational presentation mode of LLM-based news recommender systems, LLMs are likely to be influenced by users' initial interaction with a certain type of news (e.g., a particular political viewpoint), which could disproportionately influence future recommendations.

\item \textbf{Framing bias~\cite{tversky1981framing}:} The way information is presented (framed) can influence decision-making and judgments. For example, news headlines and summaries framed in a particular way can lead LLM-based news recommender systems to severely prefer these news articles.

\item \textbf{Status quo bias~\cite{samuelson1988status}:} The tendency to favor familiar content that has appeared in previous experiences. In LLM-based news recommender systems, LLM may prefer news articles that they have seen in the pre-training or fine-tuning stages, resulting in users finding it hard to view the latest news articles and reducing the diversity of information consumed.

\item \textbf{Group attribution bias~\cite{hamilton1976illusory}:} This type of bias refers to the tendency to associate specific topics or opinions with particular demographic groups. E.g., LLM-based news recommender systems may disproportionately recommend certain news topics to specific ethnic or social groups, reinforcing harmful biases and deepening societal divisions.

\end{itemize}

In conclusion, cognitive biases in LLM-based news recommender systems can have significant effects on society. These biases may trap users in information bubbles, exposing them mostly to content that reinforces their existing beliefs and preferences. Over time, this can lead to increased social polarization, reduced critical thinking, and a less informed public. The risks posed by cognitive biases in LLM-based news recommender systems highlight the need for strategies that mitigate these biases and promote more balanced and diverse news recommendations.


%% file: sections/02-mitigate.tex
\section{Mitigating Strategies}

LLM-based news recommender systems are trained on extensive human-generated datasets, which inherently reflect cognitive biases present in human society. LLM-based news recommender systems might inadvertently absorb and replicate these biases, leading to outputs that may not only reflect but also reinforce existing cognitive biases. 

To address the cognitive biases of LLM-based news recommender systems, we propose several strategies to mitigate cognitive biases through data augmentation, prompt engineering and learning algorithms aspects. Below is a detailed list of mitigating strategies:
\begin{itemize}[leftmargin=*,nosep]
\item \textbf{Synthetic data augmentation:}  LLMs can be employed to generate synthetic datasets that are carefully crafted to break correlations between cognitive biases and irrational outputs~\cite{ding2024data,lyu2024knowtuning}. First, we can construct balanced synthetic datasets that counteract the skewed cognitive bias pattern existing in human-generated datasets. Then, we train LLM-based news recommender systems on balanced datasets to reduce the influence of cognitive biases and generate more rational outputs. This strategy helps to reduce specific biases in human-generated datasets, such as group attribution bias, by ensuring that the training data represents a more diverse and equitable distribution of content.

\item \textbf{Self-debiasing via iterative refinement:} LLMs possess a self-refinement ability that can be harnessed through specially designed debiasing prompts. By iteratively refining the outputs, the model can be guided to recognize and correct biases in its recommendations. For example, in the context of LLM-based news recommender systems, LLMs can be prompted to reconsider their choices and adjust them to minimize the influence of cognitive biases. This process involves LLMs generating outputs, evaluating outputs against debiasing criteria, and revising outputs if necessary~\cite{madaan2024self}.

\item \textbf{Cognitive debiasing through human feedback:}  After an LLM generates a news recommendation, human evaluators can assess whether the output exhibits cognitive biases. If biases are detected, techniques such as direct preference optimization (DPO)~\cite{rafailov2024direct} or reinforcement learning from human feedback (RLHF)~\cite{stiennon2020learning} can be employed to adjust the model's behavior. This learning from the human feedback process helps to align the model's outputs more closely with human ethical standards and reduces the likelihood of biased recommendations in the future. By integrating human feedback into the training loop, the model learns to prioritize objectivity and fairness, thus mitigating the effects of cognitive biases.

\end{itemize}

Overall, LLM-based news recommender systems fine-tuned on extensive human-generated data are susceptible to cognitive biases. However, developing strategies such as synthetic data augmentation, self-debiasing via iterative refinement, and cognitive debiasing through human feedback can help mitigate these biases, fostering more rational and equitable outputs.  These strategies are essential for ensuring that LLMs serve as reliable components in decision-making processes for news recommendations.

%% file: sections/03-related_work.tex
\section{Related Work}
Large language models (LLMs) have become essential components of news recommender systems due to their advanced generation abilities. \citet{li2023preliminary} directly prompt ChatGPT~\cite{DBLP:conf/nips/Ouyang0JAWMZASR22} to generate news recommendations. Prompt4NR~\cite{zhang2023prompt} introduces the prompt learning paradigm to news recommendation by reframing the task of predicting user clicks on news articles as a cloze-style mask-prediction problem. PGNR~\cite{li2024prompt} transfers the personalized news recommendation task into a text-to-text generation task for LLMs, following a generative training and inference paradigm that directly generates recommendations. ONCE~\cite{liu2024once} investigates the integration of both open- and closed-source LLMs to improve news recommendation. While the deployment of LLMs into news recommender systems has led to significant improvements, previous works ignore the potential risks associated with cognitive bias in LLMs for news recommender systems.

Recent studies have shown that LLMs are affected by cognitive biases when making decisions~\cite{jones2022capturing,agrawal2023large,lin2023mind,lu2021fantastically,schmidgall2024addressing,tjuatja2023llms,lyu2023feature,ribeiro2020beyond}. \citet{jones2022capturing} identify that error patterns of GPT-3~\cite{brown2020language} resemble human cognitive biases. Similarly, \citet{agrawal2023large} discover the framing effect bias of GPT-3~\cite{brown2020language} in the clinical information extraction task. \citet{DBLP:journals/corr/abs-2308-00225} suggest that LLMs develop emergent cognitive biases after fine-tuning on extensive human-generated data. Furthermore, \citet{DBLP:journals/corr/abs-2403-00811} introduce a framework for the quantitative evaluation of cognitive biases in LLMs within a student admissions task. To the best of our knowledge, we are the first to discuss the influence of cognitive biases and mitigation strategies in LLM-based news recommender systems.

%% file: sections/04-conclusion.tex
\section{Conclusion}

In this study, we focus on exploring possible risks of cognitive biases in LLM-based recommender systems and discussing potential mitigating techniques. Given the critical role news recommender systems play in creating public opinion and influencing social discourse, any distortion brought about by cognitive biases could have far-reaching effects. To explore the risks of cognitive bias, we have introduced the general definitions of anchoring bias, framing bias, status quo bias and group attribution bias, and how they specifically affect the decision-making process of LLM-based news recommender systems. Our discussion of the cognitive biases inherent in LLMs reveals that, without careful deployment, these biases can lead to skewed and incorrect outputs in news recommendations. Our analysis of the cognitive biases in LLM-based news recommender systems shows that these biases can cause biased and inaccurate news recommendations without careful deployment of LLMs. To address these challenges, we have proposed mitigating strategies from three aspects, including data augmentation, prompt engineering and learning algorithms. 

While LLMs demonstrate the potential to enhance the personalization and relevance of news recommendations, their implementation must be carefully managed to avoid exacerbating cognitive biases. By implementing the proposed strategies, we can create more reliable and equitable news recommender systems, better meeting the diverse needs of users and contributing to a healthier news recommendation ecosystem.